\documentstyle[12pt]{article}

\def\bs{\begin{subequations}}
\def\es{\end{subequations}}

\catcode`\@=11

\newtoks\@stequation
\def\subequations{\refstepcounter{equation}
  \edef\@savedequation{\the\c@equation}%
  \@stequation=\expandafter{\theequation}
  \edef\@savedtheequation{\the\@stequation}
  \edef\oldtheequation{\theequation}%
  \setcounter{equation}{0}%
  \def\theequation{\oldtheequation\alph{equation}}}

\def\endsubequations{\setcounter{equation}{\@savedequation}%
  \@stequation=\expandafter{\@savedtheequation}%
  \edef\theequation{\the\@stequation}\global\@ignoretrue}

\catcode`\@=12

\catcode`\@=11
\def\vereq#1#2{\lower3pt\vbox{\baselineskip1.5pt \lineskip1.5pt
\ialign{$\m@th#1\hfill##\hfil$\crcr#2\crcr\sim\crcr}}}
\catcode`\@=12

\makeatletter
        \renewcommand{\theequation}{\thesection.\arabic{equation}}%
        \@addtoreset{equation}{section}%
\makeatother

\renewcommand{\thefootnote}{\fnsymbol{footnote}}

\begin{document}
\begin{titlepage}

May 4, 2006   
\begin{center}        \hfill   \\
            \hfill     \\
                                \hfill   \\

\vskip .25in

{\large \bf Further Computations of 
the He Atom Ground State \\}

\vskip 0.3in

Charles Schwartz \footnote{Email: schwartz@physics.berkeley.edu}

\vskip 0.3in

{\em Department of Physics,
     University of California\\
     Berkeley, California 94720  }
        
\end{center}

\vskip .3in

\vfill

\begin{abstract}
Recently reported computations have been extended to give ten more 
decimals of accuracy in the ground state energy of the Schrodinger 
equation for the idealized Helium atom. With the F basis - Hylleraas 
coordinates with negative powers and a logarithm of $s$ - carried to the 
fiftieth order (N = 24,099 terms) we find the eigenvalue \newline 
E = -2.90372 43770 34119 59831 11592 45194 40444 66969 25309 \ldots
\end{abstract}

\vfill

\vskip 1cm

\end{titlepage}

\renewcommand{\thefootnote}{\arabic{footnote}}
\setcounter{footnote}{0}
\renewcommand{\thepage}{\arabic{page}}
\setcounter{page}{1}

In a recently published paper \cite{me}, I reported the results of large scale 
systematic computations of the ground state eigenvalue of the 
Hamiltonian for the idealized Helium atom,
\begin{equation}
H = -\frac{1}{2} [\bigtriangledown_{1}^{2} + \bigtriangledown_{2}^{2}]
 -2/r_{1} -2/r_{2} + 1/r_{12} ,
\end{equation}
using the variational method with several basis sets formed with the 
Hylleraas coordinates,
\begin{equation}
s = r_{1} + r_{2}, \;\;\;\;\; t=r_{1} - r_{2}, \;\;\;\;\; u=r_{12} = 
|\vec{x}_{1}-\vec{x}_{2}|.
\end{equation}

The most rapid convergence was found with the ``F-Basis'',
\bs
\begin{eqnarray}
\psi = \sum C_{l,m,n}\; (1, ln s) \;e^{-ks/2}\; s^{l}\;(u/s)^{m}\;(t/s)^{n} \\
l,m = 0,1,2,3,\ldots , \;\;\; n=0,2,4,6,\ldots 
\end{eqnarray}
\es
We designate a calculation of order $\omega$ to mean a 
basis set including all terms with $l+m+n \leq \omega$. Here, the 
scale parameter is fixed at $k=2$.

The previous results went through $\omega = 37$; and now we report 
continued results through $\omega = 50$.  The table below presents 
the new results, as a continuation of Table 2 in the previous 
publication. Also shown is the Ratio of successive differences which 
is one measure of convergence rate.  Figure 1 in the previous paper 
showed a very rapid rise in the accuracy with increasing N, the number 
of basis functions used; and this new data shows no diminution of that 
rapid climb.

\vskip 1cm

\begin{tabular}{|r|r||l|c|}
\multicolumn{4}{l}{\textbf{} New Calculated Results with the F-Basis} \\
 \hline  
 $\omega$ & N  &  Energies & Ratios \\ \hline
 36 & 9499 & `` 40438 342 & 3.11 \\
 37 & 10259 & 40444 00495 & 11.1 \\ 
 38 & 11057 & 40444 51579 435 & 3.81 \\
 39 & 11897 & `` 65044 4349 & 8.65 \\
 40 & 12779 & 66593 038 & 4.84 \\
 41 & 13703 & 66913 05205  & 6.81 \\
 42 & 14671 & 66960 00893 6 & 6.18 \\
 43 & 15683 & 66967 621 & 5.43 \\
 44 & 16741 & 66969 023 & 7.66 \\
 45 & 17845 & 66969 20593  & 4.44 \\
 46 & 18997 & `` 24711 8 & 9.00 \\
 47 & 20197 & 25170 030 & 3.75 \\
 48 & 21447 & 25292 13 & 9.10 \\
 49 & 22747 & 25305 571 & 3.14 \\
 50 & 24099 & 25309 838 & \\ \hline
\end{tabular}
\vskip 1 cm 
The ditto marks `` in the table indicate that blocks of repeated 
digits have been left out, so that we can better see the new digits 
at each step.

With this data we can extrapolate to: \newline 
E* = -2.90372 43770 34119 59831 11592 45194 40444 66969 25310 5


\vskip 1cm

\begin{thebibliography}{99}

\bibitem{me} 
Charles Schwartz, ``Experiment and Theory in Computations of the He 
Atom Ground State'',  {\sl International Journal of 
Modern Physics E}\/ {\bf vol. 15,  no. 
4}, pp. 877-888 (2006). The original preprint is posted at the e-Print
arXiv: physics/0208004 ; and some further information is available at 
http://socrates.berkeley.edu/\verb+~+schwrtz/physics.html 

\end{thebibliography}
\end{document}